\documentclass[sigconf,screen]{acmart}
\usepackage{array}
\usepackage{graphicx}
\usepackage{xcolor}
\AtBeginDocument{}

\acmConference[]{}{}{}
\renewcommand\footnotetextcopyrightpermission[1]{}
\begin{document}
\fancyhead{}

\title{ACE-Bench: A Lightweight Benchmark for Evaluating Azure SDK Usage Correctness}


\author{Wenxing Zhu}
\orcid{0000-0002-8878-3286}
\email{wenxingzhu@microsoft.com}
\affiliation{%
    \institution{Microsoft}
    \city{Shanghai}
    \country{China}
}

\author{Simeng Qi}
\orcid{0009-0004-1905-7223}
\affiliation{%
    \institution{Microsoft}
    \city{Shanghai}
    \country{China}
}

\author{Junkui Chen}
\orcid{0009-0005-3989-2119}
\affiliation{%
    \institution{Microsoft}
    \city{Shanghai}
    \country{China}
}

\author{Yan Xie}
\orcid{0009-0001-4888-6977}
\affiliation{%
    \institution{Microsoft}
    \city{Shanghai}
    \country{China}
}

\author{Min Huang}
\orcid{0009-0000-9468-9301}
\affiliation{%
    \institution{Microsoft}
    \city{Shanghai}
    \country{China}
}

\author{Jingkan He}
\orcid{0009-0000-1064-5193}
\affiliation{%
    \institution{Microsoft}
    \city{Shanghai}
    \country{China}
}

\author{Xiao Wang}
\orcid{0009-0006-0613-7667}
\affiliation{%
    \institution{Microsoft}
    \city{Shanghai}
    \country{China}
}

\author{Cheng Chen}
\orcid{0009-0000-5901-427X}
\affiliation{%
    \institution{Microsoft}
    \city{Shanghai}
    \country{China}
}

\author{Sijing Meng}
\orcid{0009-0001-1554-1518}
\affiliation{%
    \institution{Microsoft}
    \city{Shanghai}
    \country{China}
}

\author{Tianqi Zhang}
\orcid{0009-0006-0549-4507}
\email{tianqi.zhang@microsoft.com}
\authornote{Corresponding author}
\affiliation{%
    \institution{Microsoft}
    \city{Shanghai}
    \country{China}
}

\renewcommand{\shortauthors}{Zhu et al.}

\begin{abstract}
    We present ACE-Bench (Azure SDK Coding Evaluation Benchmark), an execution-free benchmark that provides fast, reproducible pass/fail signals for whether large language model (LLM)-based coding agents use Azure SDKs correctly---without provisioning cloud resources or maintaining fragile end-to-end test environments.
    ACE-Bench turns official Azure SDK documentation examples into self-contained coding tasks and validates solutions with task-specific atomic criteria: deterministic regex checks that enforce required API usage patterns and reference-based LLM-judge checks that capture semantic workflow constraints.
    This design makes SDK-centric evaluation practical in day-to-day development and CI: it reduces evaluation cost, improves repeatability, and scales to new SDKs and languages as documentation evolves.
    Using a lightweight coding agent, we benchmark multiple state-of-the-art LLMs and quantify the benefit of retrieval in an MCP-enabled augmented setting, showing consistent gains from documentation access while highlighting substantial cross-model differences.
\end{abstract}





\maketitle

\section{Introduction}
Large language models (LLMs) have advanced rapidly in recent years, with frontier systems demonstrating strong performance on a wide range of programming tasks.\cite{Codet5, AlphaCode, CodeGenerationSurvey, AgentCodeGenerationSurvey, HumanInTheLoop-Software-Agents}
This progress has accelerated the adoption of LLM-based coding agents in software development: agents can interpret natural-language requirements, plan multi-step changes across a codebase, invoke tools, and iteratively refine implementations.
As agentic coding systems are integrated into real development workflows, a practical question becomes central: how do we evaluate whether a coding agent uses APIs correctly and produces code that meets task requirements?

Prior benchmarks such as HumanEval\cite{HumanEval} focus on algorithmic synthesis, while BigCodeBench\cite{BigCodeBench}, StackEval\cite{StackEval}, SWE-bench\cite{SWE-Bench} and related benchmarks\cite{SWE-PolyBench, SWE-Sharp-Bench, SWE-Multimodal} target realistic repository-level bug fixing.
These benchmarks are highly valuable, but they typically rely on executing tests.
For cloud-provider SDK tasks, execution-based evaluation is often expensive or impractical because it requires provisioning resources, managing credentials/quotas, and maintaining brittle end-to-end environments.

LLM-as-a-judge evaluation\cite{LLM-as-a-Judge,LLM-as-a-Judge-Survey} is a promising execution-free alternative.
Prior work is either reference-free\cite{SocREval,OpinSummEval,LLM-as-a-Judge}, relying on the judge model alone, or reference-based\cite{JudgeBench,JudgeLM,FActScore,freitag-etal-2021-results, Karpinska_Iyyer_2023}, anchoring judgments in trusted sources to improve reliability.

Inspired by prior work, we present ACE-Bench (Azure SDK Coding Evaluation Benchmark), an execution-free benchmark grounded in official documentation with hybrid criteria: deterministic regex checks and a reference-based LLM judge.
We use ACE-Bench to quantify both cross-model coding capability differences (RQ1) and the impact of tool augmentation via MCP-based retrieval-augmented generation (RAG)\cite{RAG} (RQ2).

In addition to being lightweight to run, ACE-Bench is designed to be low-cost to build and extend.
In our internal AI-assisted engineering workflow ("vibe coding")\cite{VibeCoding}, three engineers built the first runnable end-to-end version (documentation collection, dataset generation, and evaluation) in three days, helping validate the practical effectiveness of the Microsoft Learn MCP Tool\footnote{https://github.com/MicrosoftDocs/mcp} early in its development.

\section{ACE-Bench}
ACE-Bench is built from official SDK documentation examples.
Given cloud SDK documentation, we use an LLM to derive coding tasks and synthesize task-specific evaluation criteria that capture required API usage patterns as well as semantic constraints that target common failure modes beyond surface-level syntax.
Each task provides a natural-language prompt to the coding agent; the agent's output is then scored against a set of atomic evaluation rules, grounded in a documentation-derived reference answer.
This execution-free design avoids the cost of provisioning cloud test environments and is inexpensive to extend: adding new SDKs or languages requires only updating the documentation source, not building new runtime infrastructure.

The remainder of this section describes the dataset format (Section~\ref{sec:dataset-format}), the multi-step creation pipeline(Section~\ref{sec:dataset-creation}), and the evaluation methodology (Section~\ref{sec:evaluation-methodology}).

\begin{figure*}[t]
    \centering
    \IfFileExists{images/refined-evaluation.png}{%
        \includegraphics[width=\textwidth]{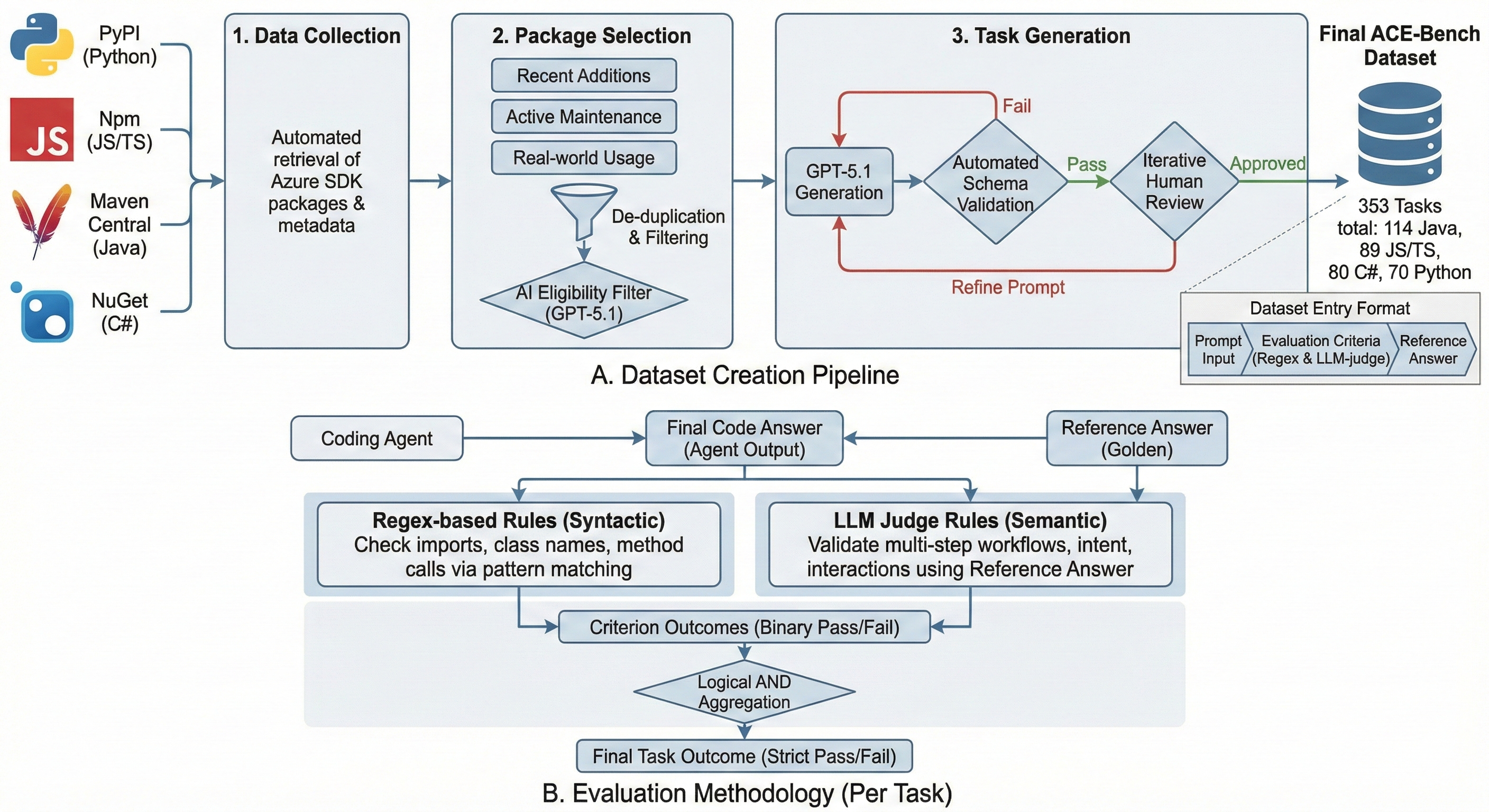}%
    }{%
        \fbox{\parbox[c][0.22\textheight][c]{0.95\textwidth}{\centering Placeholder figure: images/refined-evaluation.png}}%
    }
    \caption{Overview of the ACE-Bench workflow. (A) Dataset creation: we collect Azure SDK packages and metadata from four ecosystems, select eligible packages via de-duplication/filtering and an LLM-based eligibility check, and generate tasks with automated schema validation and iterative human review. (B) Per-task evaluation (execution-free): given a task prompt, a coding agent produces a final code answer that is checked against atomic criteria derived from official SDK documentation. Syntactic requirements are enforced deterministically via regex-based rules, while semantic workflow requirements are assessed by a reference-grounded LLM judge. A task is marked as a strict pass only if all criteria pass (logical AND).}
    \label{fig:acebench-eval}
\end{figure*}

\subsection{Dataset Format}
\label{sec:dataset-format}
Each dataset entry in ACE-Bench corresponds to a self-contained coding task (task instance) along with its validation logic.
The structure specifically isolates the input prompt from the evaluation criteria, enabling clean separation between agent execution and scoring.
In the remainder of the paper, we refer to a dataset entry as a (coding) task unless the data schema is explicitly discussed.
Effectively, each task consists of the following 3 primary components:
\begin{enumerate}
    \item \textbf{Prompt Input to the Coding Agent}: The prompt input to the coding agent. This content covers \textit{scenario context} (e.g., ``Using the Azure Blob Storage client library for Python...'') and a \textit{specific question} that defines the coding objective (e.g., ``upload a file named 'data.txt' to container 'logs'''). This separation mimics realistic developer queries where intent is often framed by a broader project context.
    \item \textbf{Evaluation Criteria}: A list of atomic assertions that define correctness (typically 3--5 criteria per entry). Each criterion specifies a validation method and what must be satisfied. We use two complementary validation types: \textit{regex-based rules} that check syntactic requirements via pattern matching (e.g., verifying specific import statements and required class or method names), and \textit{LLM judge rules} that assess semantic correctness for criteria that cannot be captured through simple pattern matching (e.g., overall code intent alignment with the task objective, multi-step workflows, negative constraints, cross-component interactions).
    \item \textbf{Reference Answer}: A documentation-grounded reference implementation extracted from official SDK guidance. It serves as the anchor for semantic-judge checks, provides the intended behavior against which an agent's output is compared. Each task has a single reference answer, which serves as the shared anchor for all semantic-judge criteria associated with that entry.
\end{enumerate}

\subsection{Dataset Creation}
\label{sec:dataset-creation}
We built the ACE-Bench dataset through a multi-stage pipeline that collects Azure SDK packages, filters them strategically, and transforms their documentation into tasks.
The process balances automation with quality control to generate hundreds of high-quality tasks across four programming languages.

\subsubsection{Data Collection}
We begin by automatically retrieving Azure SDK packages from four major repositories: PyPI\footnote{https://pypi.org/} for Python, npm\footnote{https://www.npmjs.com/} for JavaScript/TypeScript, Maven Central\footnote{https://central.sonatype.com/} for Java, and NuGet\footnote{https://www.nuget.org/} for C\#.
We implement language-specific fetchers that query each repository's API and identify Azure-related packages using organizational ownership and naming conventions.
For each SDK, we record its documentation, version metadata, and programming language.

\subsubsection{Package Selection}
With hundreds of packages available per language, we adopt a deterministic, multi-criteria selection strategy to obtain a compact yet representative subset for benchmark construction.
Concretely, we construct a candidate set by taking the union of three slices designed to capture complementary aspects of the ecosystem:
\begin{itemize}
    \item Recent additions: the newest packages ranked by first upload time (top 5).
    \item Active maintenance: the most recently released packages ranked by last release time (top 10; top 45 for Java to account for the larger and more fragmented package ecosystem).
    \item Real-world usage: the most frequently downloaded packages (top 35), using last-month downloads when available and falling back to total downloads otherwise.
\end{itemize}
We then merge these slices and de-duplicate by package name, yielding a curated set that typically contains on the order of tens of packages (SDKs) per language (up to 50 package candidates before de-duplication in most languages).
To ensure the selected packages are suitable for task generation, we additionally filter out packages with insufficient textual documentation (e.g., missing summary and description).

We then apply an LLM-based eligibility filter\footnotemark{} that evaluates each package against four criteria: presence of concrete code snippets, active maintenance status, distinct SDK identity (not meta-packages), and reasonable compatibility requirements (e.g., excluding Python 2-only packages).
Packages failing any criterion are excluded to avoid wasted generation costs.

\subsubsection{Task Generation}
For each eligible package, an LLM\footnotemark[\value{footnote}] generates up to three dataset entries conforming to the schema in Section~\ref{sec:dataset-format}.
\footnotetext{We use GPT-5.1 (\url{https://openai.com/index/gpt-5-1/}) for both eligibility filtering and task generation.}
We further implement rigorous validation to ensure all generated tasks meet required quality standards.
Every task undergoes automated schema validation to verify data integrity: required fields are present and correctly typed, metadata is complete, and evaluation rules are properly specified.
Tasks failing validation are rejected, and the system tracks success rates for each package to identify cases where documentation may be insufficient for automatic task generation.

We also use an iterative human review process to refine the generation prompt.
In each iteration, we inspect a stratified sample of the generated tasks (20\% with a minimum of 30 tasks per iteration), ensuring coverage across languages and oversampling packages with low generation success rates.
Reviewers follow a consistent checklist: (i) the task statement is SDK-specific and unambiguous, (ii) the reference (golden) answer matches official documentation, and (iii) the evaluation criteria meaningfully validate the intended behavior rather than superficial patterns.
For each sampled task, reviewers assign an outcome (accept / revise / reject) and record the primary issue type.
Typical revisions include clarifying underspecified prompts, correcting reference answers to align with the cited documentation, and tightening or relaxing regex rules to reduce false positives/negatives on intended variants.
Issues are categorized by severity (e.g., critical misalignment with documentation, under-specified tasks, or ineffective criteria); disagreements are resolved via discussion among reviewers, and any systematic failure triggers prompt/template updates followed by regeneration.
We iterate until two consecutive review rounds surface no critical issues in the sampled set, indicating that the prompt consistently produces tasks that meet our quality bar.

After completing data collection, package selection, and task generation, the current ACE-Bench release contains 353 tasks spanning Java (114), JavaScript/TypeScript (89), C\# (80), and Python (70).

\subsection{Evaluation Methodology}
\label{sec:evaluation-methodology}

For each task in the dataset, we instruct the coding agent to generate a final code snippet that achieves the specified objective described in the prompt.
We evaluate only this final output, ignoring any intermediate reasoning steps and tool-call traces, against a set of atomic criteria (defined separately from the prompt) grounded in a documentation-based reference answer.
We use two complementary validation types:

\begin{itemize}
    \item \textbf{Regex-based rules} validate whether the coding agent has applied the correct SDK packages, key class names, and method calls. Each criterion specifies a regular expression pattern and uses pattern matching to verify required API usage signatures (e.g., specific import statements and client and method identifiers), yielding a binary score: match or no match.
    \item \textbf{LLM-judge rules} evaluate semantic correctness for requirements that are hard to express as patterns. Each criterion specifies a semantic check (e.g., whether the code satisfies a required intent, or correctly follows a multi-step workflow). For each LLM-judge criterion, the judge compares the generated code against the intent captured by the prompt and the reference answer, and returns a binary decision: pass or fail.
\end{itemize}

After evaluating all regex-based criteria and LLM-judge criteria, we obtain a set of Boolean outcomes.
We compute a strict pass/fail decision as the logical AND over all criteria.
Figure~\ref{fig:acebench-eval} also summarizes the evaluation workflow, where regex-based criteria enforce concrete API-usage signatures (e.g., required imports and client and method identifiers), while LLM-judge criteria validate semantic intent and multi-step workflows, reducing false positives from overly permissive judging.

\section{Experiments}
This section evaluates whether ACE-Bench provides a reliable, \\execution-free signal for Azure SDK usage correctness, and whether that signal is sensitive enough to reveal meaningful capability differences across both models and agent configurations.
Our experiments are designed around controlled comparisons that isolate the impact of information availability while keeping the agent interface and evaluation protocol fixed.

We study the following research questions.
\begin{itemize}
    \item \textbf{RQ1 (Cross-model sensitivity).} Can ACE-Bench distinguish coding capability differences among different foundation models under the same agent framework?
    \item \textbf{RQ2 (Tooling sensitivity).} For the same coding agent, can ACE-Bench measure performance differences when the agent is equipped with different levels of external tool access?
\end{itemize}

\subsection{Experiment Setup}
We implement a lightweight coding agent using the mcp-use SDK\footnote{https://pypi.org/project/mcp-use/}, which provides a unified interface for optionally invoking MCP tools during problem solving.
To evaluate cross-model differences (RQ1), we run the same agent with different LLM backends across multiple model families (e.g., OpenAI GPT-series\footnote{https://platform.openai.com/docs/models}, Anthropic Claude-series\footnote{https://platform.claude.com/docs/en/about-claude/models/overview}, and Grok-series models\footnote{https://docs.x.ai/docs/models}).
We select representative models spanning a range of capability and cost tiers.
To isolate the model's intrinsic knowledge, we run a \textit{non-augmented} setting that disables all external tools and documentation access, and the agent answers solely from its pre-trained knowledge.
To evaluate tool sensitivity (RQ2), we additionally run an \textit{augmented} setting where the agent is equipped with the Microsoft Learn MCP server.
This tool allows the agent to retrieve up-to-date information from Microsoft Learn\footnote{https://learn.microsoft.com/} documentation (including Azure SDK documentation) via MCP before producing the final code. We evaluate $353$ ACE-Bench tasks under both \textit{non-augmented} and \textit{augmented} settings. Across all runs, we evaluate only the final code output and discard intermediate reasoning and tool-call traces, ensuring that the scoring reflects end-to-end coding correctness rather than the verbosity of the agent.

\subsection{Results and Discussion}
We report results for 11 models in Table~\ref{tab:exp_results_passrate}.
Following dataset format described in Section~\ref{sec:dataset-format} and our evaluation methodology in Section~\ref{sec:evaluation-methodology}, we report a \textit{strict pass rate}, defined as the fraction of tasks for which the agent satisfies \emph{all} atomic criteria.
Table~\ref{tab:exp_results_passrate} summarizes strict pass rates for each model under both settings.

\begin{table}
    \caption{Strict pass rate by model. \textit{Non-augmented} denotes the non-augmented setting without external documentation/tool access; \textit{Augmented} denotes the setting with Microsoft Learn MCP documentation access enabled. Values are reported as point estimates with 95\% Wilson score confidence intervals computed over per-task strict-pass outcomes; we display intervals in the table as estimate $\pm$ half-width (computed from the Wilson interval bounds). $\Delta$ is the absolute improvement (percentage points).}
    \label{tab:exp_results_passrate}
    \begin{tabular}{>{\raggedright\arraybackslash}p{0.28\linewidth}rrr}
        \toprule
        Model                     & Non-aug.(\%)   & Aug.(\%)       & $\Delta$ (pp) \\
        \midrule

        claude-opus-4.1           & 34.3 $\pm$ 4.9 & 53.6 $\pm$ 5.3 & 19.3          \\
        claude-haiku-4.5          & 25.8 $\pm$ 4.5 & 58.0 $\pm$ 5.1 & 32.2          \\
        claude-sonnet-4.5         & 34.3 $\pm$ 4.9 & 63.7 $\pm$ 5.1 & 29.5          \\
        claude-opus-4.5           & 39.4 $\pm$ 5.1 & 65.3 $\pm$ 5.0 & 26.0          \\
        gpt-4.1                   & 27.8 $\pm$ 4.7 & 51.1 $\pm$ 5.2 & 23.4          \\
        gpt-5-mini                & 26.9 $\pm$ 4.6 & 49.6 $\pm$ 5.3 & 22.6          \\
        gpt-5                     & 34.3 $\pm$ 4.9 & 53.5 $\pm$ 5.4 & 19.2          \\
        gpt-5.1                   & 32.0 $\pm$ 4.8 & 63.2 $\pm$ 5.0 & 31.2          \\
        grok-4                    & 31.7 $\pm$ 4.8 & 68.7 $\pm$ 4.8 & 36.9          \\
        grok-4-fast-non-reasoning & 14.7 $\pm$ 3.7 & 50.9 $\pm$ 5.3 & 36.2          \\
        grok-code-fast-1          & 24.4 $\pm$ 4.5 & 58.5 $\pm$ 5.2 & 34.1          \\
        \bottomrule
    \end{tabular}
\end{table}

\textbf{RQ1 (Cross-model sensitivity).}
ACE-Bench clearly distinguishes coding performance across model backends under the same agent framework.
In the non-augmented setting, strict pass rates range from $14.7\%$ to $39.4\%$ across the evaluated models.
In the augmented setting, the spread remains substantial ($49.6\%$ to $68.7\%$), indicating that even with the same tool access and prompts, models differ meaningfully in their ability to consistently satisfy the full set of SDK-usage constraints.
Table~\ref{tab:exp_results_passrate} reports Wilson-score 95\% confidence intervals for these pass rates.
The between-model spread is large relative to the per-model Wilson intervals in both settings, indicating clear separation that is not attributable to minor sampling fluctuations.
Overall, these results suggest that ACE-Bench provides a sensitive, execution-free signal for cross-model comparison.

\begin{center}
    {\setlength{\fboxrule}{1pt}\setlength{\fboxsep}{6pt}\fbox{\begin{minipage}{\dimexpr\columnwidth-2\fboxsep-2\fboxrule\relax}
                \textbf{Answer to RQ1:} Under a fixed agent, ACE-Bench reliably separates model backends, demonstrating strong cross-model sensitivity for SDK-centric coding correctness.
            \end{minipage}}}
\end{center}

\textbf{RQ2 (Tooling sensitivity).}
In the augmented setting, enabling MCP-based documentation retrieval leads to consistent improvements across all tested models.
Averaged over the 11 models in Table~\ref{tab:exp_results_passrate}, strict pass rate increases from $29.6\%$ (non-augmented) to $57.8\%$ (augmented), an average gain of $+28.2$ percentage points.
The largest improvement is observed for grok-4 ($+36.9$ points), while the smallest improvement is about $+19$ points (e.g., gpt-5 at $+19.2$ points).
Across models, these improvements are consistently positive and sizeable, while the Wilson intervals in Table~\ref{tab:exp_results_passrate} indicate relatively tight uncertainty around each per-setting strict pass rate.
Overall, the consistent positive deltas support that ACE-Bench is sensitive to retrieval/tool augmentation and can quantify the benefit of MCP-enabled documentation access for SDK-centric coding tasks.

\begin{center}
    {\setlength{\fboxrule}{1pt}\setlength{\fboxsep}{6pt}\fbox{\begin{minipage}{\dimexpr\columnwidth-2\fboxsep-2\fboxrule\relax}
                \textbf{Answer to RQ2:} Enabling documentation/tool access consistently improves performance, confirming ACE-Bench's sensitivity to tool augmentation.
            \end{minipage}}}
\end{center}

\subsubsection{Effect Size Summary}
Concretely, for each task we compute a \textit{criterion satisfaction score} as the mean of its atomic criterion outcomes, and for each model and setting we report this score averaged over tasks.
We then compute per-model deltas (augmented minus non-augmented) and report the unweighted mean delta across the 11 models.
Under this protocol, MCP augmentation increases the mean per-task \textit{criterion satisfaction score} by $+0.18$ on average, and increases strict pass rate by about $+0.28$ on average.
Together with the cross-model spread in Table~\ref{tab:exp_results_passrate}, these systematic improvements indicate that ACE-Bench provides a sensitive, execution-free signal for measuring coding capability differences across model backends and agent tool configurations.

\subsubsection{Case Study}

\begin{figure}[t]
    \centering
    \setlength{\fboxsep}{6pt}

    \begin{minipage}[t]{\linewidth}
        \centering
        \textbf{Error-Catching Regex Rule}\par
        \vspace{4pt}
        \fbox{\begin{minipage}[t]{0.98\linewidth}
                {\ttfamily\footnotesize\raggedright
                    new\textbackslash s+MySQLManagementFlexibleServerClient\textbackslash s*\textbackslash(\par
                }
            \end{minipage}}
    \end{minipage}

    \vspace{6pt}

    \begin{minipage}[t]{\linewidth}
        \centering
        \textbf{Incorrect Agent Output}\par
        \vspace{4pt}
        \fbox{\begin{minipage}[t]{0.98\linewidth}
                {\ttfamily\footnotesize\raggedright
                    import \{ DefaultAzureCredential \} from "@azure/identity";\par
                    import \{ \textcolor{red}{MySQLManagementClient} \} from "\textcolor{red}{@azure/arm-mysql-flexible}";\par
                    \par
                    const subscriptionId = "YOUR\_SUBSCRIPTION\_ID";\par
                    const credential = new DefaultAzureCredential();\par
                    \par
                    const client = new \textcolor{red}{MySQLManagementClient}(credential, subscriptionId);\par
                }
            \end{minipage}}
    \end{minipage}

    \vspace{6pt}

    \begin{minipage}[t]{\linewidth}
        \centering
        \textbf{Reference Answer}\par
        \vspace{4pt}
        \fbox{\begin{minipage}[t]{0.98\linewidth}
                {\ttfamily\footnotesize\raggedright
                    import \{ \textcolor{green!60!black}{MySQLManagementFlexibleServerClient} \} from "\textcolor{green!60!black}{@azure/arm-mysql-flexible}";\par
                    import \{ DefaultAzureCredential \} from "@azure/identity";\par
                    \par
                    const subscriptionId = "YOUR\_SUBSCRIPTION\_ID";\par
                    const credential = new DefaultAzureCredential();\par
                    \par
                    const client = new \textcolor{green!60!black}{MySQLManagementFlexibleServerClient}(credential, subscriptionId);\par
                }
            \end{minipage}}
    \end{minipage}

    \caption{Import--client mismatch in Azure MySQL Flexible Server management (JavaScript). The incorrect snippet imports from the flexible-server management package but instantiates a generic MySQL management client; ACE-Bench's regex rules reject this deterministically.}
    \label{fig:mysql-flex-import-mismatch}
\end{figure}

In a JavaScript task that asks the agent to create an Azure MySQL Flexible Server management client using Azure AD default credentials, the model-produced answer can appear superficially correct (ES module imports, DefaultAzureCredential, and a placeholder subscription ID), and a pure LLM-as-a-judge check is likely to mark it as pass.
However, the answer uses the wrong management client and package: it instantiates the generic MySQL management client while importing it from the flexible-server management package.
In the Azure JavaScript SDK, the generic MySQL management client is provided by the MySQL management package, whereas managing MySQL Flexible Server requires the flexible-server management client from the flexible-server package.
This is a representative failure mode where an LLM produces plausible-looking code by reusing familiar class names across closely related SDK packages. Figure~\ref{fig:mysql-flex-import-mismatch} shows a concrete instance of this import--client mismatch. ACE-Bench's regex-based rules catch this deterministically by enforcing the correct import signature, thereby preventing false positives that can arise from judge hallucinations or overly permissive semantic grading.

\subsubsection{Threats to Validity}
ACE-Bench is derived from official documentation examples, which biases tasks toward documented best practices and may under-represent long-tail production edge cases.
SDKs and docs evolve; even with retrieval, reference answers and regex rules can become stale.
Our execution-free scoring combines deterministic regex checks with a reference-grounded LLM judge, but semantic outcomes can still be sensitive to judge choice and prompting.
We mitigate these risks by grounding judge checks in documentation-derived references, enforcing concrete API signatures via regex, and applying the human review gate in Section~2 to audit doc-faithfulness and criterion effectiveness.

\section{Conclusion and Future Work}
ACE-Bench is a documentation-grounded, execution-free benchmark for evaluating Azure SDK usage correctness.
Our experiments show that ACE-Bench is sensitive to cross-model differences under a fixed agent framework (RQ1) and consistently measures gains from MCP-based documentation retrieval (RQ2), supporting its use as a practical signal for SDK-centric coding.

Future work includes expanding coverage beyond Azure to other major cloud-provider SDKs and strengthening semantic evaluation (e.g., more structured judging and calibration), as well as exploring richer agent settings and complementary lightweight static signals.


\bibliographystyle{ACM-Reference-Format}
\bibliography{reference}










\end{document}